\documentclass[english,conference]{IEEEtran}
\usepackage[T1]{fontenc}
\usepackage[latin9]{inputenc}
\usepackage{array}
\usepackage{units}
\usepackage{amsmath}
\usepackage{graphicx}
\usepackage{amssymb}

\makeatletter

\providecommand{\tabularnewline}{\\}
\newtheorem{definitn}{Definition}
\newtheorem{conject}{Conjecture}

\usepackage{flushend}

\makeatother

\usepackage{babel}

\begin{document}

\title{Unimodular Lattices for the Gaussian Wiretap Channel}

\author{\IEEEauthorblockN{Jean-Claude Belfiore and Patrick Solé}
\IEEEauthorblockA{Telecom ParisTech, LTCI UMR 5141\\ 
46, rue Barrault\\75013 Paris, France\\ 
Email: \{jean-claude.belfiore, patrick.sole\}@telecom-paristech.fr}}
\maketitle
\begin{abstract}
In \cite{sec-gain} the authors introduced a lattice invariant called
{}``Secrecy Gain'' which measures the confusion experienced by a
passive eavesdropper on the Gaussian Wiretap Channel. We study, here,
the behavior of this invariant for unimodular lattices by using tools
from Modular Forms and show that, for some families of unimodular
lattices, indexed by the dimension, the secrecy gain exponentially
goes to infinity with the dimension. 
\end{abstract}

\section{Introduction}

The wiretap channel was introduced by Wyner \cite{Wyner-II} as a
discrete memoryless broadcast channel where the sender, Alice, transmits
confidential messages to a legal receiver Bob, in the presence of
an eavesdropper Eve. Wyner defined the perfect secrecy capacity as
the maximum amount of information that Alice can send to Bob while
insuring that Eve gets a negligeable amount of information. He also
described a generic coding strategy known as coset coding. While coset
coding has been used in many coding scenarios (for ex. \cite{ZSE-02,PR-05}),
Wyner used it to encode both data and random bits to confuse the eavesdropper.
The question of determining the secrecy capacity of many classes of
channels has been addressed extensively recently, yielding a plethora
of information theoretical results on secrecy capacity.

There is a sharp contrast with the situation of wiretap code designs,
where very little is known. The most exploited approach to get practical
codes so far has been to use LDPC codes (for example \cite{TDCMM-07}
for binary erasure and symmetric channels, \cite{KHMBK-09} for Gaussian
channels with binary inputs). We also note that wiretap II codes have
been extended to more general settings such as network coding in \cite{ES-07}.
Finally, lattice codes for Gaussian channels have been considered
from an information theoretical point of view in \cite{HY-09}.

In \cite{sec-gain}, a design criterion for constructing explicit
lattice codes, has been proposed, based on the analysis of Eve's correct
decision probability. This design criterion relies on a new lattice
invariant called {}``secrecy gain'' based on theta series. In this
paper, we analyze the secrecy gain for unimodular lattices. 

The paper is organized as follows. In Section \ref{sec:The-secrecy-gain}
we recall the definition of the secrecy gain and give its value for
some extremal even unimodular lattices. An asymptotic analysis is
then performed in section \ref{sec:Asymptotic-Analysis} for even
unimodular lattices and we prove that the secrecy gains of some of
these lattices grow up to infinity with the dimension. Finally we
come back to the Gaussian wiretap channel in section \ref{sec:Back-to-the}
to show that unimodular lattices only define one operating point for
the system. Some other types of lattices should be studied in the
future for other operating points.

\section{Notations and previous results\label{sec:Notations-and-previous}}

\subsection{Notations and system model}

We use, in this paper, the same system model and the same notations
as \cite{sec-gain}. In \cite{sec-gain}, the secrecy gain has been
defined and some examples have been given. Here we analyze more deeply
this parameter for even unimodular lattices and give the asymptotic
behavior of this secrecy gain when the dimension of the lattices grows
to infinity. %
\begin{figure}[ht]

\noindent \begin{centering}
\includegraphics[width=0.8\columnwidth]{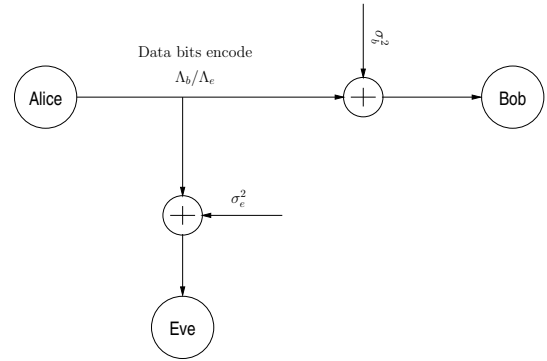}
\par\end{centering}

\caption{\label{fig:The-Gaussian-Wiretap}The Gaussian Wiretap Channel}

\end{figure}
Figure \ref{fig:The-Gaussian-Wiretap} gives the model considered
in this paper where Alice wants to send data to Bob on a Gaussian
channel whose noise variance is given by $\sigma_{b}^{2}$. Eve is
the eavesdropper trying to intercept data through another Gaussian
channel whose noise variance is $\sigma_{e}^{2}$. In order to have
a positive secrecy capacity, we will assume that $\sigma_{e}^{2}>\sigma_{b}^{2}$.
Bits are transmitted by Alice at a rate equal to $R=R_{s}+R_{r}$
where $R_{s}$ is the secrecy rate of this transmission and $R_{r}$
is the rate of pseudo-random bits. Indeed, we use Wyner's generic
coding strategy \cite{Wyner-I}. We give the remaining parameters,
\begin{itemize}
\item $\Lambda_{b}$ is the fine lattice (used to minimize Bob's probability
of error)
\item $\Lambda_{e}$ is the coarse lattice (used to minimize Eve's probability
of correct decision)
\item $n$ is the dimension of both lattices
\item $\mathcal{V}\left(\Lambda_{b}\right)$ (resp. $\mathcal{V}\left(\Lambda_{e}\right)$)
is the fundamental parallelotope of $\Lambda_{b}$ (resp. $\Lambda_{e}$)
\item $\mbox{Vol}\left(\mathcal{P}\right)$ is the volume of $\mathcal{P}$
\end{itemize}
Data bits label cosets in $\Lambda_{b}/\Lambda_{e}$ while pseudo-random
bits label points of $\Lambda_{e}$. The reader can refer to \cite{sec-gain}
for a more detailed description of the coding scheme. Still according
to \cite{sec-gain}, and under the assumption of a moderate to high
secrecy rate, the expression of the probability of correct decision
at the eavesdropper can be expressed as \begin{equation}
P_{c,e}\simeq\left(\frac{1}{\sqrt{2\pi}\sigma_{e}}\right)^{n}\mbox{Vol}\left(\mathcal{V}\left(\Lambda_{b}\right)\right)\sum_{\mathbf{r}\in\Lambda_{e}}e^{-\frac{\left\Vert \mathbf{r}\right\Vert ^{2}}{2\sigma_{e}^{2}}}.\label{eq:correct-eve}\end{equation}

In eq. (\ref{eq:correct-eve}), we recognize the theta series of lattice
$\Lambda_{e}$.

\subsection{Theta series of a lattice}
\begin{definitn}
\label{def:theta}Let $\Lambda$ be a Euclidean lattice, then the
theta series of $\Lambda$ is \cite{CS-98} \begin{equation}
\Theta_{\Lambda}(z)\triangleq\sum_{\boldsymbol{x}\in\Lambda}q^{\left\Vert \boldsymbol{x}\right\Vert ^{2}},q=e^{i\pi z},\mathrm{Im}(z)>0\label{eq:theta-def}\end{equation}

Some exceptional lattices have theta series that can be expressed
as functions of the Jacobi theta functions $\vartheta_{i}(q)$, $i=2,3,4$
with \begin{align*}
\vartheta_{2}(q) & =\sum_{n=-\infty}^{+\infty}q^{\left(n+\frac{1}{2}\right)^{2}}\\
\vartheta_{3}(q) & =\sum_{n=-\infty}^{+\infty}q^{n^{2}}\\
\vartheta_{4}(q) & =\sum_{n=-\infty}^{+\infty}\left(-1\right)^{n}q^{n^{2}}\end{align*}

For instance, table \ref{tab:Theta-series} gives the theta series
of some exceptional lattices. %
\begin{table}[h]
\noindent \begin{centering}
\begin{tabular}{|c|c|}
\hline 
Lattice $\Lambda$ & Theta series $\Theta_{\Lambda}$\tabularnewline
\hline
\hline 
Cubic lattice $\mathbb{Z}^{n}$ & $\vartheta_{3}^{n}$\tabularnewline
\hline 
$D_{n}$ & $\frac{1}{2}\left(\vartheta_{3}^{n}+\vartheta_{4}^{n}\right)$\tabularnewline
\hline 
Gosset lattice $E_{8}$ & $\frac{1}{2}\left(\vartheta_{2}^{8}+\vartheta_{3}^{8}+\vartheta_{4}^{8}\right)$\tabularnewline
\hline
\end{tabular}
\par\end{centering}

~

\caption{\label{tab:Theta-series}Theta series of some lattices}

\end{table}

\end{definitn}

\subsection{Minimization of the theta series}

One problem that arises naturally when studying theta series is the
following. In eq. (\ref{eq:correct-eve}), set $y=iz$ and restrict
to real values of $y$. We are now interested in studying \[
\Theta_{\Lambda}(y)=\sum_{\boldsymbol{x}\in\Lambda}q^{\left\Vert \boldsymbol{x}\right\Vert ^{2}},q=e^{-\pi y},y>0.\]
Equation (\ref{eq:correct-eve}), giving Eve's probability of correct
decision, can be written as \begin{equation}
P_{c,e}\simeq\left(\frac{1}{\sqrt{2\pi}\sigma_{e}}\right)^{n}\mbox{Vol}\left(\mathcal{V}\left(\Lambda_{b}\right)\right)\Theta_{\Lambda_{e}}\left(\frac{1}{2\pi\sigma_{e}^{2}}\right)\label{eq:correct-theta}\end{equation}
So, for a given dimension $n$, the problem to solve is to find a
lattice $\Lambda^{\mathsf{opt}}$ that minimizes $\Theta_{\Lambda}(y)$
for a given value of $y$ in order to minimize expression (\ref{eq:correct-theta}).

\section{The secrecy gain\label{sec:The-secrecy-gain}}

\subsection{Definitions}

We recall here some definitions given in \cite{sec-gain}. 

We remark that, if we do not use any specific coarse lattice $\Lambda_{e}$,
we can assume that $\Lambda_{e}$ is equal to a scaled version of
$\mathbb{Z}^{n}$ with same volume as $\Lambda_{e}$. Consequently,
for a lattice $\Lambda$, it is natural to define the secrecy function.
For a lattice with unitary volume, we have
\begin{definitn}
Let $\Lambda$ be an $n-$dimensional lattice with unitary volume.
The \textit{secrecy function} of $\Lambda$ is 
\end{definitn}
\[
\Xi_{\Lambda}(y)\triangleq\frac{\Theta_{\mathbb{Z}^{n}}(y)}{\Theta_{\Lambda}(y)}=\frac{\vartheta_{3}(y)^{n}}{\Theta_{\Lambda}(y)}\]
defined for $y>0$. 

Then of course, as we want to minimize the expression of Eve's probability
of correct decision in eq. (\ref{eq:correct-theta}), we are interested
in the maximum value of the secrecy function. So, we define the secrecy
gain, 
\begin{definitn}
The \textit{secrecy gain} of an $n-$dimensional lattice $\Lambda$
is \[
\chi_{\Lambda}\triangleq\sup_{y>0}\Xi_{\Lambda}(y)\]

\end{definitn}

\subsection{The secrecy gain of unimodular lattices}

Theta series are difficult to analyze. Nevertheless, for some lattices,
these functions have nice properties. It is the case of even unimodular
lattices whose theta series are modular forms with integer weight.
We mainly restrict this paper to the study of even unimodular lattices
and will use tools from modular forms.

\subsubsection{Definitions and formulas}

We recall the definition of an integral lattice \cite{CS-98}, 
\begin{definitn}
A lattice $\Lambda$ is integral if its Gram matrix has entries in
$\mathbb{Z}$. Note that an integral lattice has the property\[
\Lambda\subseteq\Lambda^{\star}\subseteq\frac{1}{\mbox{Vol}\left(\mathcal{V}\left(\Lambda\right)\right)^{2}}\Lambda\]

\end{definitn}
From this definition, we can now define unimodular lattices, 
\begin{definitn}
A lattice $\Lambda$ is unimodular if 
\begin{enumerate}
\item $\Lambda$ is integral 
\item $\Lambda$ is equal to its dual
\end{enumerate}
Note that a unimodular lattice has fundamental volume equal to $1$. 
\end{definitn}
Let $\Lambda^{\star}$ be the dual lattice of the $n-$dimensional
lattice $\Lambda$. Then Jacobi's formula \cite{CS-98} gives the
theta series of $\Lambda^{\star}$ as a function of the theta series
of $\Lambda$, \begin{equation}
\Theta_{\Lambda^{\star}}(y)=\mbox{Vol}\left(\mathcal{V}\left(\Lambda\right)\right)y^{-\frac{n}{2}}\Theta_{\Lambda}\left(\frac{1}{y}\right)\label{eq:jacobi}\end{equation}

If $\Lambda$ is unimodular, then using (\ref{eq:jacobi}), we deduce
\[
\Theta_{\Lambda}(y)=\Theta_{\Lambda^{\star}}(y)=y^{-\frac{n}{2}}\Theta_{\Lambda}\left(\frac{1}{y}\right).\]
So, since $\mathbb{Z}^{n}$ itself is unimodular, the secrecy function
of $\Lambda$ has the property, \[
\Xi_{\Lambda}(y)=\Xi_{\Lambda}\left(\frac{1}{y}\right).\]
 If we express $y$ in decibel (in our case, $y=\tfrac{1}{2\pi\sigma_{e}^{2}}$
and is related to Eve's signal to noise ratio), then the secrecy function
becomes an even function. 
\begin{conject}
\label{con:secrecy-unimodular}The secrecy gain of unimodular lattices
is achieved by the secrecy function at $y=1$. 
\end{conject}
Using conjecture \ref{con:secrecy-unimodular} in what follows, we
can evaluate the secrecy gain of unimodular lattices as \[
\chi_{\Lambda}=\Xi_{\Lambda}(1)\]

\paragraph*{Some formulas}

Some formulas are useful to calculate the secrecy gain of unimodular
lattices. The most important ones, found in \cite{jacobi-theta},
are \begin{eqnarray}
\vartheta_{2}\left(e^{-\pi}\right) & = & \vartheta_{4}\left(e^{-\pi}\right)\nonumber \\
\vartheta_{3}\left(e^{-\pi}\right) & = & \sqrt[4]{2}\vartheta_{4}\left(e^{-\pi}\right)\label{eq:formulas}\end{eqnarray}

\subsubsection{Secrecy gain of some exceptional unimodular lattices}

\paragraph{Gosset Lattice $E_{8}$}

$E_{8}$ is unimodular even. From table \ref{tab:Theta-series} and
eq. (\ref{eq:formulas}), we get \begin{eqnarray*}
\frac{1}{\Xi_{E_{8}}(1)} & = & \frac{\tfrac{1}{2}\left(\vartheta_{2}(e^{-\pi})^{8}+\vartheta_{3}(e^{-\pi})^{8}+\vartheta_{4}(e^{-\pi})^{8}\right)}{\vartheta_{3}(e^{-\pi})^{8}}\\
 & = & \tfrac{1}{2}\left(1+\frac{1}{4}+\frac{1}{4}\right)\\
 & = & \frac{3}{4}\end{eqnarray*}
We deduce, then, the secrecy gain of $E_{8}$, \[
\boxed{\chi_{E_{8}}=\Xi_{E_{8}}(1)=\frac{4}{3}=1.33333}\]

\begin{figure}[ht]
\noindent \begin{centering}
\includegraphics[width=0.8\columnwidth]{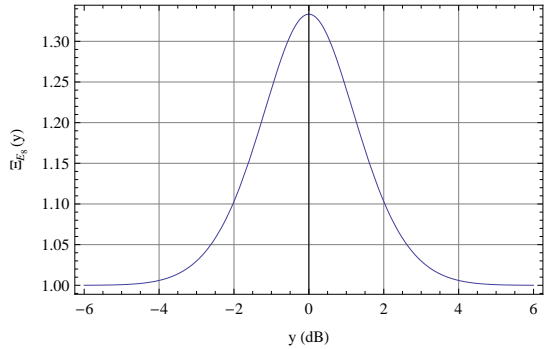}
\par\end{centering}

\caption{\label{fig:Secrecy-function-E8}Secrecy function of $E_{8}$}

\end{figure}

As an illustration, figure \ref{fig:Secrecy-function-E8}%
 gives the secrecy function of $E_{8}$.

\paragraph{Leech Lattice $\Lambda_{24}$}

$\Lambda_{24}$ is also unimodular even. From table \ref{tab:Theta-series},
we get (with simplified notations) \begin{eqnarray*}
\frac{1}{\Xi_{\Lambda_{24}}(1)} & = & \frac{\frac{1}{8}\left(\vartheta_{2}^{8}+\vartheta_{3}^{8}+\vartheta_{4}^{8}\right)^{3}-\frac{45}{16}\vartheta_{2}^{8}\vartheta_{3}^{8}\vartheta_{4}^{8}}{\vartheta_{3}^{24}}\\
 & = & \frac{27}{2^{6}}-\frac{45}{2^{8}}\\
 & = & \frac{63}{256}\end{eqnarray*}
We deduce, then, the secrecy gain of $\Lambda_{24}$, \[
\boxed{\chi_{\Lambda_{24}}=\Xi_{\Lambda_{24}}(1)=\frac{256}{63}=4.0635}\]

\begin{figure}[ht]
\noindent \begin{centering}
\includegraphics[width=0.8\columnwidth]{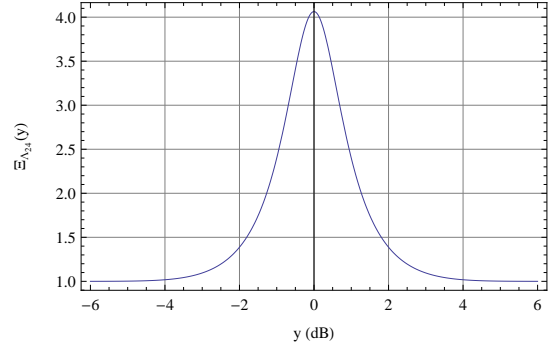}
\par\end{centering}

\caption{\label{fig:Secrecy-function-Leech}Secrecy function of $\Lambda_{24}$}

\end{figure}

As an illustration, figure \ref{fig:Secrecy-function-Leech}%
 gives the secrecy function of $\Lambda_{24}$.

\subsection{Higher dimension unimodular extremal lattices}

$E_{8}$ and $\Lambda_{24}$ are extremal even unimodular lattices
in dimensions $8$ and $24$ respectively \cite{CS-98}. Extremal
means that their minimum distance is maximal for a given dimension
\cite{CS-98}. We can give same type of results for extremal even
unimodular lattices of higher dimensions. For instance, we can derive
the secrecy functions and secrecy gains of extremal even unimodular
lattices in dimensions $32$, $48$, and $72$ using derivations of
\cite{Skoruppa}. The same can be done in dimension $80$ by solving
a linear system \cite{Ebeling}. Please note that, until now, nobody
knows if an extremal lattice in dimension $72$ exists. Results are
summarized in table \ref{tab:Theta-series-extremal}. %
\begin{table}[h]
\noindent \begin{centering}
\begin{tabular}{|c|c|c|}
\hline 
Dimension & Lattice $\Lambda$ & $\Theta_{\Lambda}$\tabularnewline
\hline
\hline 
$8$ & $E_{8}$ & $E_{4}$\tabularnewline
\hline 
$24$ & $\Lambda_{24}$ & $E_{4}^{3}-720\Delta$\tabularnewline
\hline 
$32$ & $BW_{32}$ & $E_{4}^{4}-960E_{4}\Delta$\tabularnewline
\hline 
$48$ & $P_{48}$ & $E_{4}^{6}-1440E_{4}^{3}\Delta+125280\Delta^{2}$\tabularnewline
\hline 
$72$ & $L_{72}$ & $E_{4}^{9}-2160E_{4}^{6}\Delta+965520E_{4}^{3}\Delta^{2}-27302400\Delta^{3}$\tabularnewline
\hline 
$80$ & $L_{80}$ & $E_{4}^{10}-2400E_{4}^{7}\Delta+1360800E_{4}^{4}\Delta^{2}-103488000E_{4}\Delta^{3}$\tabularnewline
\hline
\end{tabular}
\par\end{centering}

\caption{\label{tab:Theta-series-extremal}Theta series of extremal lattices}

\end{table}
 Here we introduce the function \[
\Delta(q)=\frac{E_{4}^{3}(q)-E_{6}^{2}(q)}{12^{3}}\]
where $E_{k}$ are the Eisenstein series \cite{Ebeling} defined as\begin{align}
E_{k}(q) & =1+\frac{2}{\zeta\left(1-k\right)}\sum_{m=1}^{+\infty}m^{k-1}\frac{q^{m}}{1-q^{m}}\label{eq:Eisenstein}\\
 & =1-\frac{2k}{B_{k}}\sum_{m=1}^{+\infty}m^{k-1}\frac{q^{m}}{1-q^{m}}\nonumber \end{align}
where $B_{k}$ are the Bernouilli numbers \cite{Serre-1} and $\zeta(s)$
is the Riemann zeta function \[
\zeta(s)=\sum_{k=1}^{\infty}\frac{1}{k^{s}}.\]
 Relations with Jacobi functions are (in symbolic notation) \[
\begin{cases}
E_{4} & =\frac{1}{2}\left(\vartheta_{2}^{8}+\vartheta_{3}^{8}+\vartheta_{4}^{8}\right)\\
\Delta & =\frac{1}{256}\vartheta_{2}^{8}\vartheta_{3}^{8}\vartheta_{4}^{8}\end{cases}\]

\begin{figure*}
\noindent \begin{centering}
\includegraphics[width=0.4\textwidth]{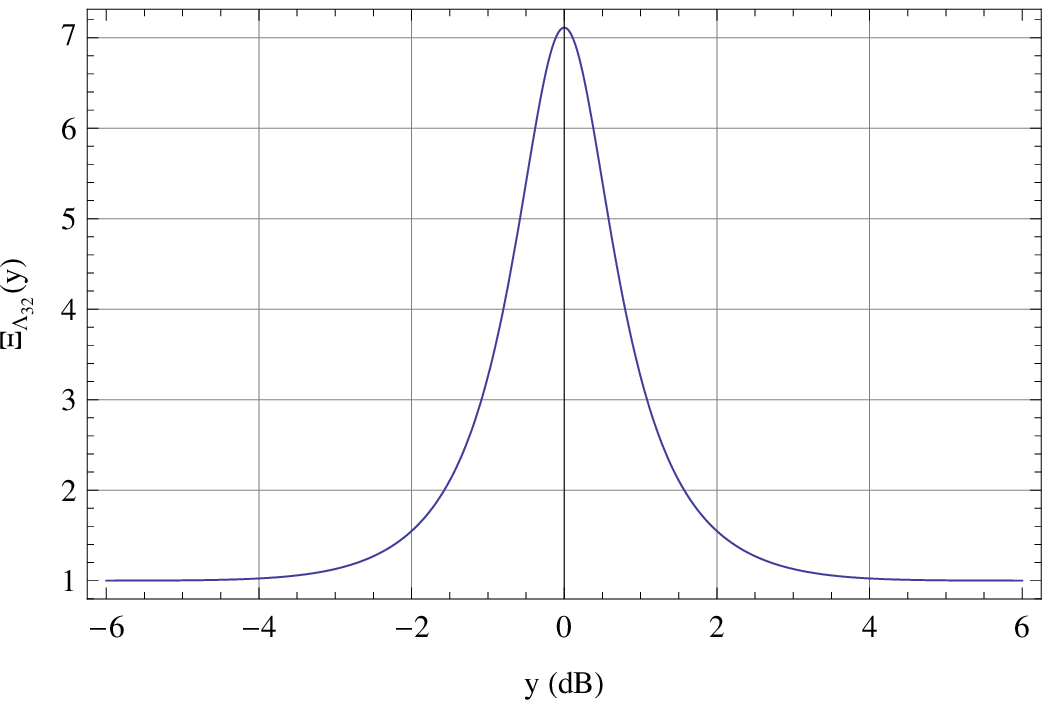}\quad{}\includegraphics[width=0.4\textwidth]{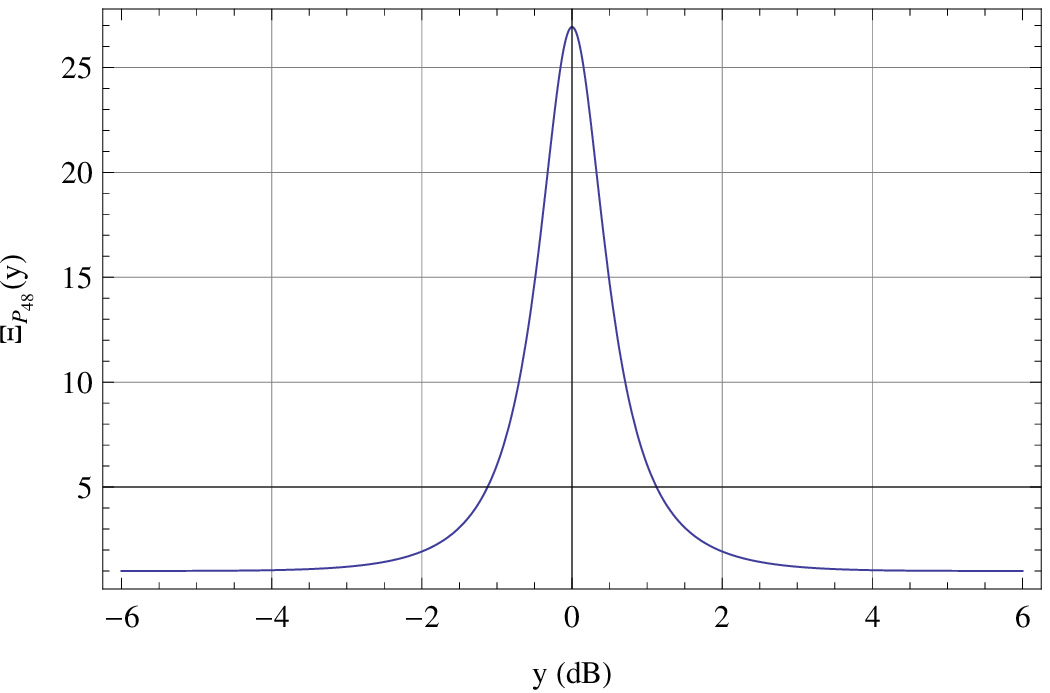}\bigskip{}

\par\end{centering}

\noindent \begin{centering}
\includegraphics[width=0.4\textwidth]{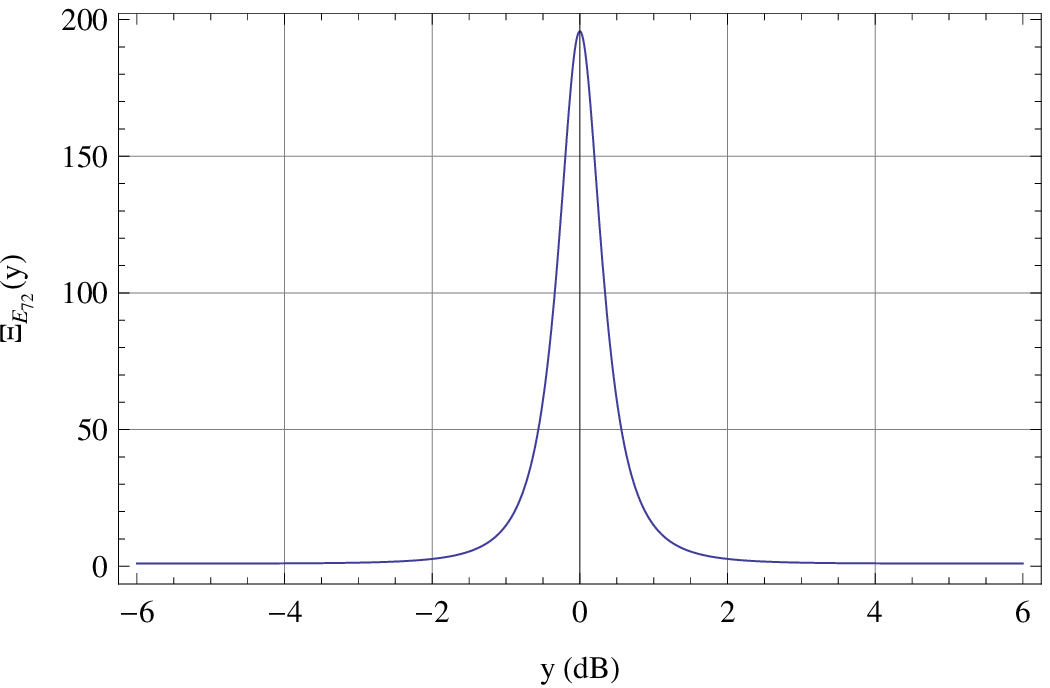}\quad{}\includegraphics[width=0.4\textwidth]{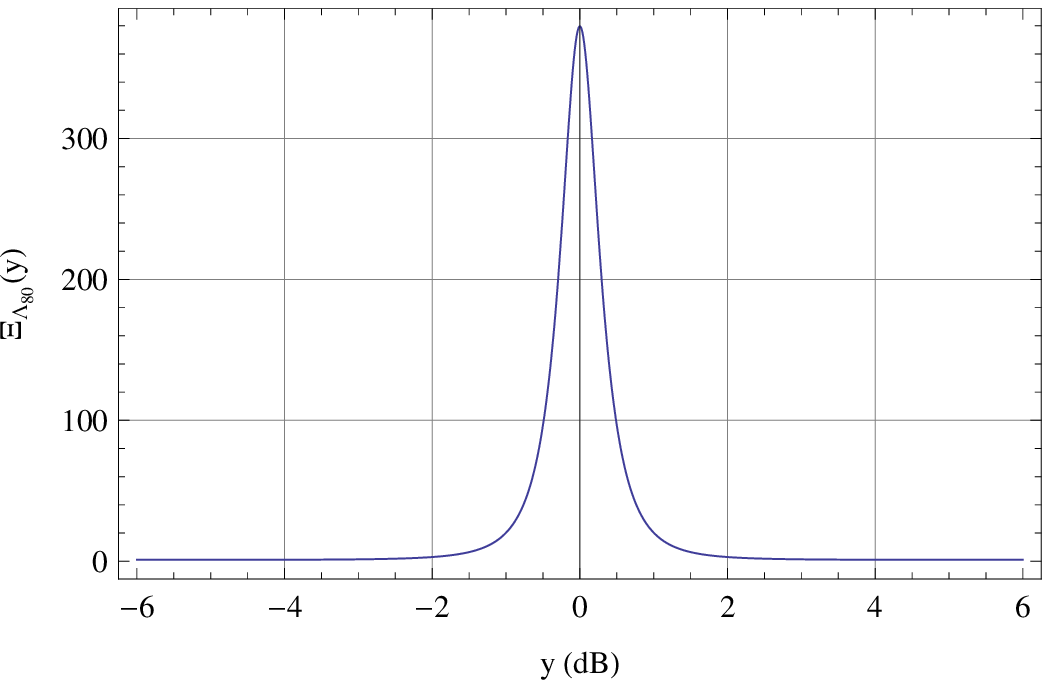}
\par\end{centering}

\caption{\label{fig:Secrecy-functions}Secrecy functions of extremal lattices
in dimensions 32, 48, 72 and 80}

\end{figure*}
and give rise to the expressions of theta series evaluated below.

\subsubsection{Barnes Wall lattice $BW_{32}$}

In dimension $32$, Barnes-Wall lattice $BW_{32}$ is an extremal
lattice. We have \begin{eqnarray*}
\Theta_{BW_{32}} & = & \frac{1}{16}\left(\vartheta_{2}^{8}+\vartheta_{3}^{8}+\vartheta_{4}^{8}\right)\left[\left(\vartheta_{2}^{8}+\vartheta_{3}^{8}+\vartheta_{4}^{8}\right)^{3}\right.\\
 &  & \left.-30\cdot\vartheta_{2}^{8}\cdot\vartheta_{3}^{8}\cdot\vartheta_{4}^{8}\right]\end{eqnarray*}
so, \begin{align*}
\frac{1}{\Xi_{BW_{32}}(1)} & =\frac{1}{16}\left(1+\frac{1}{2}\right)\left[\left(1+\frac{1}{2}\right)^{3}-30\cdot\frac{1}{16}\right]\\
 & =\frac{9}{64}.\end{align*}
Hence, \[
\chi_{BW_{32}}=\frac{64}{9}\simeq7.11\]

\subsubsection{Lattice $P_{48p(q)}$}

There are two different extremal even unimodular lattices in dimension
$48$, $P_{48p}$ and $P_{48q}$ \cite[Chap. 5]{CS-98}, having, of
course the same theta series, \begin{eqnarray*}
\Theta_{P_{48}} & = & \frac{1}{2048}\left[3915\vartheta_{2}^{16}\vartheta_{3}^{16}\vartheta_{4}^{16}\right.\\
 &  & -1440\vartheta_{2}^{8}\vartheta_{3}^{8}\vartheta_{4}^{8}\left(\vartheta_{2}^{8}+\vartheta_{3}^{8}+\vartheta_{4}^{8}\right)^{3}\\
 &  & \left.+32\left(\vartheta_{2}^{8}+\vartheta_{3}^{8}+\vartheta_{4}^{8}\right)^{6}\right]\end{eqnarray*}
giving \begin{align*}
\frac{1}{\Xi_{P_{48}}(1)} & =\frac{1}{2048}\left[\frac{3915}{256}-\frac{1440}{16}\left(1+\frac{1}{2}\right)^{3}+32\left(1+\frac{1}{2}\right)^{6}\right]\\
 & =\frac{19467}{524288}.\end{align*}
Hence, \[
\chi_{P_{48}}=\frac{524288}{19467}\simeq26.93\]

\subsubsection{Dimensions $72$ and $80$}

In the same way, from table \ref{tab:Theta-series-extremal}, we can
compute the secrecy gain for an extremal unimodular even lattice in
dimension $72$ and $80$. Note that two examples of such lattices
in dimension $80$ have been given in \cite{extremal-80}. We have
\begin{eqnarray*}
\chi_{\Lambda_{72}} & = & \frac{134217728}{685881}\simeq195.69\\
\chi_{\Lambda_{80}} & = & \frac{536870912}{1414413}\simeq379.57\end{eqnarray*}

\begin{table}[h]
\noindent \begin{centering}
\begin{tabular}{|c|c|c|c|c|c|c|}
\hline 
Dimension & 8 & 24 & 32 & 48 & 72 & 80\tabularnewline
\hline
\hline 
Secrecy gain & $1.3$ & $4.1$ & $7.11$ & $26.9$ & $195.7$ & $380$\tabularnewline
\hline
\end{tabular}
\par\end{centering}

\caption{\label{tab:Secrecy-gains-extremal}Secrecy gains of extremal lattices }

\end{table}
Table \ref{tab:Secrecy-gains-extremal} summarizes all these results.

\section{Asymptotic Analysis\label{sec:Asymptotic-Analysis}}

We propose, here to find a lower bound of the best secrecy gain as
a function of the dimension $n$ , and deduce some asymptotic results
(when $n$ is large enough). For a fixed dimension $n$, we compute
bounds on the theta series of an optimal unimodular lattice. By optimal,
we mean a lattice which maximizes the secrecy gain. We will use the
Siegel-Weil formula to compute these bounds.

\subsection{A Siegel-Weil formula for theta series of even unimodular lattices}

Let $n\equiv0\:(\mbox{mod }8)$, $\Omega_{n}$ be the set of all inequivalent
even unimodular $n-$dimensional lattices. Let $k=\nicefrac{n}{2}$.
Then, one has \cite{Serre-1} \[
\sum_{\Lambda\in\Omega_{n}}\frac{\Theta_{\Lambda}(q)}{\left|\mathrm{Aut}(\Lambda)\right|}=M_{n}\cdot E_{k}(q)\]
where \[
M_{n}=\sum_{\Lambda\in\Omega_{n}}\frac{1}{\left|\mathrm{Aut}(\Lambda)\right|}\]
and $E_{k}(q)$ is the Eisenstein series with weight $k$ even whose
expression is given in eq. (\ref{eq:Eisenstein}).

Let $\Theta_{\min}^{(n)}=\min_{\Lambda\in\Omega_{n}}\Theta_{\Lambda}$.
Then \[
\Theta_{\min}^{(n)}M_{n}\leq\sum_{\Lambda\in\Omega_{n}}\frac{\Theta_{\Lambda}}{\left|\mathrm{Aut}(\Lambda)\right|}=M_{n}E_{k}\]
giving rise to \[
\Theta_{\min}^{(n)}\leq E_{k}.\]
Define \[
\chi_{n}\triangleq\max_{\Lambda\in\Omega_{n}}\chi_{\Lambda}=\frac{\vartheta_{3}^{n}\left(e^{-\pi}\right)}{\Theta_{\min}^{(n)}\left(e^{-\pi}\right)}\]
then we get, \[
\boxed{\chi_{n}\geq\frac{\vartheta_{3}^{n}\left(e^{-\pi}\right)}{E_{k}\left(e^{-2\pi}\right)}}\]

\subsection{Limit of $E_{k}$}

Assume $q$ to be a real number $0<q<1$. We have \[
E_{k}(q)=1+\frac{2k}{\left|B_{k}\right|}\sum_{m=1}^{+\infty}m^{k-1}\frac{q^{m}}{1-q^{m}}\]
Replacing $q$ by $e^{-2\pi}$ gives \[
E_{k}\left(e^{-2\pi}\right)=1+\frac{2k}{\left|B_{k}\right|}\sum_{m=1}^{+\infty}\frac{m^{k-1}}{e^{2\pi m}-1}\]
which converges (very quickly) to $2$ when $k$ is a multiple of
$4$ that tends to infinity. Moreover, according to \cite{jacobi-theta},
we have \[
\vartheta_{3}\left(e^{-\pi}\right)=\frac{\pi^{\frac{1}{4}}}{\Gamma\left(\frac{3}{4}\right)}\simeq1.086>1\]
so, \begin{equation}
\boxed{\chi_{n}\gtrsim\frac{1}{2}\left(\frac{\pi^{\frac{1}{4}}}{\Gamma\left(\frac{3}{4}\right)}\right)^{n}\simeq\frac{1.086^{n}}{2}}\label{eq:asymptotic-secrecy}\end{equation}
which tends exponentially to infinity. %
\begin{figure}[ht]
\noindent \begin{centering}
\includegraphics[width=0.8\columnwidth]{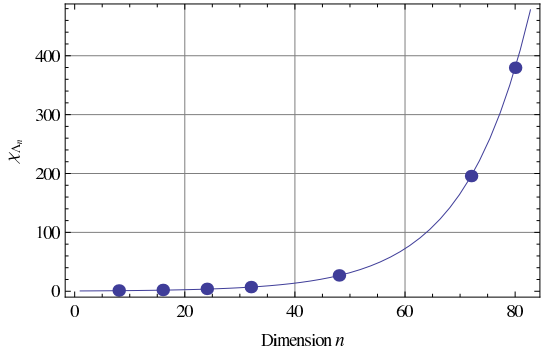}
\par\end{centering}

\caption{\label{fig:Secrecy-gain}Lower bound of the minimal secrecy gain as
a function of $n$ from Siegel-Weil formula. Points correspond to
extremal lattices.}

\end{figure}
Figure \ref{fig:Secrecy-gain} gives the asymptotic expression of
the secrecy gain as a function of the dimension $n$, as well as points
corresponding to extremal lattices in dimensions $8,16,24,32,48,72$
and $80$.

\subsection{Consequences}

We proved that there exists a family of even unimodular lattices whose
secrecy gains exponentially grows up with the dimension, which means
that Eve's probability of correct decision exponentially tends to
$0$. But as we can remark in figure \ref{fig:Secrecy-functions},
around its maximum, the secrecy function becomes sharper and sharper
when $n$ grows up, which means that, for high dimensions, the communication
system absolutely has to operate at $y=1$. We show now, in section
\ref{sec:Back-to-the}, the meaning of this constraint in terms of
the communication system as well as the way of doing the same with
$y\neq1$.

\section{Back to the wiretap channel\label{sec:Back-to-the}}

We are interested, in this section, in how the secrecy gain is related
to the parameters of the Gaussian channel, through the proposed lattice
coset construction.

\subsection{Operating point}

For unimodular lattices, the secrecy gain is obtained as the value
of the secrecy function at point $y=1$. From eq. (\ref{eq:correct-theta}),
it means \[
\sigma_{e}^{2}=\frac{1}{2\pi}.\]

Conjecture \ref{con:secrecy-unimodular} says that the secrecy gain
is achieved by the secrecy function at $y=1$ if the lattice is unimodular.
If the lattice is not unimodular, then the secrecy gain can be achieved
for another value of $y$. %
\begin{figure}[ht]
\noindent \begin{centering}
\includegraphics[width=0.8\columnwidth]{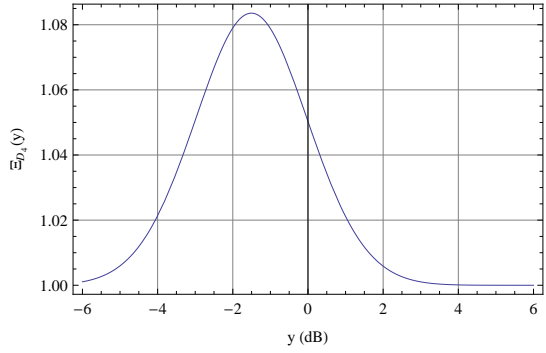}
\par\end{centering}

\caption{\label{fig:Secrecy-function-D4}Secrecy function of the checkerboard
lattice $D_{4}$}

\end{figure}
 In figure \ref{fig:Secrecy-function-D4}, we see that the secrecy
gain is achieved for $y=-1.5\mbox{ dB}$ ($y=\nicefrac{1}{\sqrt{2}}$)
and not $0\mbox{ dB}$ ($y=1$). We call \emph{operating point}, the
value $y_{\mathsf{o.p.}}$ for which \[
\chi=\Xi\left(y_{\mathsf{o.p.}}\right)\]

\subsection{Fundamental volume of $\Lambda_{e}$}

The secrecy function is defined for a fundamental volume of $\Lambda_{e}$
equal to $1$. We now relate the secrecy function to system parameters.
Let $R$ be the value of the total rate (information bits $+$ pseudo-random
bits) transmitted by Alice, per (complex) channel use and $R_{s}$
be the value of the information (secret) rate. 

We now establish a correspondence between this parameter $y$ and
the physical parameters of the channel. 

Fundamental volume of $\Lambda_{e}$ is \[
\mbox{Vol}\left(\mathcal{V}\left(\Lambda_{e}\right)\right)=2^{\frac{nR_{\mathsf{s}}}{2}}\mbox{Vol}\left(\mathcal{V}\left(\Lambda_{b}\right)\right)\]
where $R_{\mathsf{s}}$ is the rate at which Alice sends the secret
information bits to Bob. The operating point of the system (for instance
volume equal to $1$ for a unimodular lattice) corresponds, of course,
to a normalized case $y_{\mathsf{norm}}$. In practice, we should
work with a scaled lattice. The operating point of this scaled lattice
is then, \begin{eqnarray*}
y & = & y_{\mathsf{norm}}\mbox{Vol}\left(\mathcal{V}\left(\Lambda_{e}\right)\right)^{\frac{2}{n}}\\
 & = & 2^{R_{\mathsf{s}}}y_{\mathsf{norm}}\mbox{Vol}\left(\mathcal{V}\left(\Lambda_{b}\right)\right)^{\frac{2}{n}}\end{eqnarray*}
Now, the energy, per channel use, of the signal sent by Alice is \[
E_{s}=2^{R}\mbox{Vol}\left(\mathcal{V}\left(\Lambda_{b}\right)\right)^{\frac{2}{n}}\]
where $R$ is the global rate of the communication (secret bits +
pseudo-random bits). Hence, we get\[
y=2^{R_{\mathsf{s}}}y_{\mathsf{norm}}E_{s}2^{-R}=2^{-\left(R-R_{\mathsf{s}}\right)}E_{s}y_{\mathsf{norm}}.\]
 As $y_{\mathsf{norm}}=\frac{1}{2\pi\sigma_{e}^{2}}$, we get\begin{equation}
\boxed{y=\frac{2^{-\left(R-R_{\mathsf{s}}\right)}E_{s}}{2\pi\sigma_{e}^{2}}=\frac{2^{-\left(R-R_{\mathsf{s}}\right)}}{2\pi}\gamma_{e}}\label{eq:operating-point}\end{equation}
where $\gamma_{e}$ is Eve's signal to noise ratio. If we use a unimodular
lattice, then the operating point is $y=1$ which corresponds to a
secrecy rate \begin{equation}
R_{s}=R-\log_{2}\frac{2\pi}{\gamma_{e}}\label{eq:Rs-unimod}\end{equation}
Other secrecy rates require to study other families of lattices such
as modular lattices \cite{Quebbemann}, for instance.

\section{Conclusion}

The secrecy gain introduced in \cite{sec-gain} is a new lattice invariant
that measures how much confusion the eavesdropper will experience.
This parameter is based on the value of the theta series of lattice
$\Lambda_{e}$ at some point that depends on the lattice itself. We
can analyze how secrecy gain behaves, when dimension grows up, if
$\Lambda_{e}$ is an even unimodular lattice. In that case, its theta
series is a modular form with integer weight and very efficient tools
can be used to analyze its behavior. We have shown that an even unimodular
lattice with minimal theta series has a secrecy gain which exponentially
goes to infinity when dimension $n$ goes to infinity. But this only
corresponds to values of Eve's $\mathsf{SNR}$ equal to \[
\gamma_{e}=\pi2^{R-R_{s}+1}\]
where $R$ is the total bit rate and $R_{s}$ is the secrecy rate.
For other values of $\gamma_{e}$, other families of lattices have
to be considered such as $\ell-$modular lattices or their duals.
But this requires to find another Siegel-Weil formula for these types
of lattices.

\end{document}